\documentclass[12pt,a4paper]{article}

\usepackage[english]{babel}
\usepackage{psfrag,graphicx}
\usepackage{amsmath,amssymb,bbm,stmaryrd,amsthm}

\newtheorem{theorem}{Theorem}
\newtheorem{proposition}[theorem]{Proposition}
\newtheorem{lemma}[theorem]{Lemma}

\DeclareMathOperator*{\Ai}{Ai}
\DeclareMathOperator*{\Tr}{Tr}
\newcommand{\Id}{\mathbbm{1}}
\newcommand{\dx}{{\rm d}}
\newcommand{\px}{{\partial}}
\newcommand{\Rp}{{\mathbbm{R}_+}}
\newcommand{\R}{\mathbbm{R}}
\newcommand{\Z}{\mathbbm{Z}}
\newcommand{\Pb}{\mathbbm{P}}

\title{A determinantal formula for the GOE Tracy-Widom distribution}
\author{Patrik L.\ Ferrari and Herbert Spohn \\[6pt]
{\normalsize Technische Universit\"at M\"unchen}\\
{\normalsize Zentrum Mathematik and Physik Department}\\
{\normalsize e-mails: ferrari@ma.tum.de, spohn@ma.tum.de}}
\date{4th May 2005}
\begin{document}
\maketitle

\begin{abstract}
Investigating the long time asymptotics of the totally asymmetric simple exclusion process, Sasamoto obtains rather indirectly a formula for the GOE Tracy-Widom distribution. We establish that his novel formula indeed agrees with more standard expressions.
\end{abstract}

\section{Introduction}
The Gaussian orthogonal ensemble (GOE) of random matrices is a probability distribution on the set of $N\times N$ real symmetric matrices defined through
\begin{equation}
Z^{-1} e^{-\Tr(H^2)/2N}\dx H.
\end{equation}
$Z$ is the normalization constant and $\dx H=\prod_{1\leq i \leq j \leq N}\dx H_{i,j}$. The induced statistics of eigenvalues can be studied through the method of Pfaffians. Of particular interest for us is the statistics of the largest eigenvalue, $E_1$. As proved by Tracy and Widom~\cite{TW96}, the limit
\begin{equation}
\lim_{N\to\infty}\Pb\big(E_1\leq 2N+s N^{1/3}\big) = F_1(s)
\end{equation}
exists, $\Pb$ being our generic symbol for probability of the event in parenthesis. $F_1$ is called the GOE Tracy-Widom distribution function. Following~\cite{For00} it can be expressed in terms of a Fredholm determinant in the Hilbert space $L^2(\R)$ as follows,
\begin{equation}\label{I2}
F_1(s)^2=\det\big(\Id-P_s (K+|g\rangle\langle f|) P_s\big),
\end{equation}
where $K$ is the Airy kernel defined through
\begin{eqnarray}
K(x,y) &=& \int_\Rp\dx\lambda \Ai(x+\lambda)\Ai(y+\lambda), \nonumber \\
  g(x) &=& \Ai(x), \\
  f(y) &=& 1-\int_\Rp \dx \lambda \Ai(y+\lambda), \nonumber
\end{eqnarray}
and $P_s$ is the projection onto the interval $[s,\infty)$.

The GOE Tracy-Widom distribution $F_1(s)$ turns up also in the theory of one-dimensional growth process in the KPZ universality class, KPZ standing for Kardar-Parisi-Zhang~\cite{KPZ86}. Let us denote the height profile of the growth process at time $t$ by $h(x,t)$, either $x\in \R$ or $x\in\Z$. One then starts the growth process with flat initial conditions, meaning $h(x,0)=0$, and considers the height above the origin $x=0$ at growth time $t$. For large $t$ it is expected that
\begin{equation}\label{a1}
h(0,t)=c_1 t + c_2 t^{1/3} \xi_1.
\end{equation}
Here $c_1$ and $c_2$ are constants depending on the details of the model and $\xi_1$ is a random amplitude with
\begin{equation}\label{a2}
\Pb(\xi_1\leq s)=F_1(s).
\end{equation}

For the polynuclear growth (PNG) model the height $h(0,t)$ is related to the length of the longest increasing subsequence of symmetrized random permutations~\cite{PS00}, for which Baik and Rains~\cite{BR99} indeed prove the asymptotics (\ref{a1}), (\ref{a2}), see~\cite{Fer04} for further developments along this line. Very recently Sasamoto~\cite{Sas05} succeeds in proving the corresponding result for the totally asymmetric simple exclusion process (TASEP). If $\eta_j(t)$ denotes the occupation variable at $j\in\Z$ at time $t$, then the TASEP height is given by
\begin{equation}\label{1.11}
h(j,t)=
 \begin{cases}
 2N_t +\sum^j_{i=1}(1-2\eta_i(t)) & \textrm{for }j\geq 1,\\
 2N_t & \textrm{for }j=0,\\
 2N_t -\sum^0_{i=j+1}(1-2\eta_i(t)) & \textrm{for }j\leq -1,
 \end{cases}
\end{equation}
with $N_t$ denoting the number of particles which passed through the bond $(0,1)$ up to time $t$. The flat initial condition for the TASEP is $\ldots 0\,1\,0\,1\,0\,1\,\ldots$. For technical reasons Sasamoto takes instead $\ldots 0\,1\,0\,1\,0\,0\,0\,0\,0\ldots$ and studies the asymptotics of $h(-3t/2,t)$ for large $t$ with the result
\begin{equation}\label{eqTS}
h(-3t/2,t)= \tfrac12 t+\tfrac12 t^{1/3} \xi_{\rm SA}.
\end{equation}
The distribution function of the random amplitude $\xi_{\rm SA}$ is
\begin{equation}
\Pb(\xi_{\rm SA}\leq s)=F_{\rm SA}(s)
\end{equation}
with
\begin{equation}\label{e8}
F_{\rm SA}(s)=\det(\Id-P_s A P_s).
\end{equation}
Here $A$ has the kernel $A(x,y)=\frac12\Ai((x+y)/2)$ and, as before, the Fredholm determinant is in $L^2(\R)$.

The universality hypothesis for one-dimensional growth processes claims that in the scaling limit, up to model-dependent coefficients, the asymptotic distributions are identical. In particular, since (\ref{a1}) is proved for PNG, the TASEP with flat initial conditions should have the same limit distribution function, to say
\begin{equation}\label{I1}
F_{\rm SA}(s)=F_1(s).
\end{equation}
Our contribution provides a proof for (\ref{I1}).

\section{The identity}
As written above, the $s$-dependence sits in the projection $P_s$. It will turn out to be more convenient to transfer the $s$-dependence into the integral kernel. From now on the determinants are understood as Fredholm determinants in $L^2(\Rp)$ with scalar product $\langle\cdot,\cdot\rangle$. Thus, whenever we write an integral kernel like $A(x,y)$, the arguments are understood as $x\geq 0$ and $y\geq 0$.

Let us define the operator $B(s)$ with kernel
\begin{equation}
B(s)(x,y)=\Ai(x+y+s).
\end{equation}
By~\cite{TW94} $\|B(s)^2\|<1$ and clearly $B(s)$ is symmetric. Thus also $\|B(s)\|<1$ for all $s$. $B(s)$ is trace class with both positive and negative eigenvalues. Shifting the arguments in (\ref{e8}) by $s$, one notes that
\begin{equation}
F_{\rm SA}(s)=\det(\Id-B(s)).
\end{equation}
Applying the same operation to (\ref{I2}) yields
\begin{equation}
F_1(s)^2=\det\big(\Id-B(s)^2-|g\rangle\langle f|\big)
\end{equation}
with
\begin{eqnarray}
  g(x) &=& \Ai(x+s)=(B(s)\delta)(x), \\
  f(y) &=& 1-\int_\Rp \dx \lambda \Ai(y+\lambda+s)=((\Id-B(s))1)(y). \nonumber
\end{eqnarray}
Here $\delta$ is the $\delta$-function at $x=0$ and $1$ denotes the function $1(x)=1$ for all $x\geq 0$. $\delta$ and $1$ are not in $L^2(\Rp)$. Since the kernel of $B(s)$ is continuous and has super-exponential decay, the action of $B(s)$ is unambiguous.

\begin{proposition} With the above definitions we have 
\begin{equation}\label{eq2}
\det(\Id-B(s))=F_1(s).
\end{equation}
\end{proposition}
\begin{proof}
For simplicity we suppress the explicit $s$-dependence of $B$. We rewrite
\begin{eqnarray}
F_1(s)^2&=&\det\big((\Id-B)(\Id+B-|B\delta\rangle\langle1|)\big) \nonumber \\
&=&\det(\Id-B)\det(\Id+B)\big(1-\langle\delta,B(\Id+B)^{-1}1\rangle\big) \nonumber \\
&=&\det(\Id-B)\det(\Id+B)\langle\delta,(\Id+B)^{-1}1\rangle
\end{eqnarray}
since $1=\langle \delta,1\rangle$. Thus we have to prove that
\begin{equation}\label{eq3}
\det(\Id-B)=\det(\Id+B)\langle\delta,(\Id+B)^{-1} 1\rangle.
\end{equation}
Taking the logarithm on both sides,
\begin{equation}\label{eq4}
\ln\det(\Id-B)=\ln\det(\Id+B)+\ln \langle\delta,(\Id+B)^{-1} 1\rangle,
\end{equation}
and differentiating it with respect to $s$ results in
\begin{equation}\label{eq5}
-\Tr\big((\Id-B)^{-1}\frac{\px}{\px s}B)\big)= \Tr\big((\Id+B)^{-1}\frac{\px}{\px s}B)\big) +\frac{\frac{\px}{\px s}\langle\delta,(\Id+B)^{-1} 1\rangle}{\langle\delta,(\Id+B)^{-1} 1\rangle}
\end{equation}
where we used
\begin{equation}
\frac{\dx}{\dx s} \ln(\det(T)) = \Tr\big(T^{-1}\frac{\px}{\px s}T\big).
\end{equation}
Since $B(s)\to 0$ as $s\to\infty$, the integration constant for (\ref{eq5}) vanishes and we have to establish that 
\begin{equation}\label{eqFin}
-2\Tr\big((\Id-B^2)^{-1}\frac{\px}{\px s}B)\big)= \frac{\frac{\px}{\px s}\langle\delta,(\Id+B)^{-1} 1\rangle}{\langle\delta,(\Id+B)^{-1} 1\rangle}.
\end{equation}
Define the operator $D=\frac{\dx}{\dx x}$. Then using the cyclicity of the trace and Lemma~\ref{Lemma1},
\begin{eqnarray}\label{eq6}
-2\Tr\big((\Id-B^2)^{-1}\frac{\px}{\px s}B)\big)& =& -2\Tr\big((\Id-B^2)^{-1}DB)\big) \nonumber \\
&=& \langle \delta, (\Id-B^2)^{-1} B \delta\rangle.
\end{eqnarray}
Using Lemma~\ref{Lemma2} and $D1=0$, one obtains
\begin{equation}\label{eq7}
\langle \delta,\frac{\px}{\px s} (\Id+B)^{-1} 1\rangle = \langle \delta, (\Id-B^2)^{-1} B\delta\rangle \langle\delta , (\Id+B)^{-1} 1\rangle.
\end{equation}
Thus (\ref{eqFin}) follows from (\ref{eq6}) and (\ref{eq7}).
\end{proof}

\begin{lemma}\label{Lemma1}
Let $A$ be a symmetric, trace class operator with smooth kernel and let $D=\frac{\dx}{\dx x}$. Then
\begin{equation}\label{eqLem1}
2\Tr(DA)=-\langle \delta , A \delta\rangle
\end{equation}
where $DA$ is the operator with kernel $\frac{\px}{\px x} A(x,y)$.
\end{lemma}
\begin{proof}
The claim follows from spectral representation of $A$ and the identity
\begin{equation}
\int_\Rp \dx x f'(x)f(x)=-f(0)f(0)-\int_\Rp \dx x f(x)f'(x).
\end{equation}
\end{proof}

\begin{lemma}\label{Lemma2}
It holds
\begin{equation}
\frac{\px}{\px s}(\Id+B)^{-1}=(\Id-B^2)^{-1} B D + (\Id-B^2)^{-1} |B \delta \rangle \langle \delta (\Id+B)^{-1}|.
\end{equation}
\end{lemma}
\begin{proof}
First notice that $\frac{\px}{\px s} B\equiv \dot{B}=DB$. For any test function $f$,
\begin{eqnarray}
(\dot{B}f)(x)&=&\int_\Rp \dx y \partial_y \Ai(x+y+s) f(y)\nonumber \\
&=& -\Ai(x+s) f(0)-\int_\Rp\dx y \Ai(x+y+s) f'(y).
\end{eqnarray}
Thus, using the notation $P=|B\delta\rangle\langle \delta |$, one has
\begin{equation}\label{eq10}
DB=-BD-P.
\end{equation}
Since $\|B\|<1$, we can expand $\frac{\px}{\px s}(\Id+B)^{-1}$ in a power series and get
\begin{equation}\label{eq12}
\frac{\px}{\px s} (\Id+B)^{-1}=\sum_{n\geq 1}(-1)^n \frac{\px}{\px s} B^n = \sum_{n\geq 1}(-1)^n \sum_{k=0}^{n-1} B^k D B^{n-k}.
\end{equation}
Using recursively (\ref{eq10}) we obtain
\begin{eqnarray}\label{eq11}
\sum_{k=0}^{n-1} B^k D B^{n-k} &=& -\frac{1-(-1)^n}{2} B^n D +\sum_{j=0}^{n-1}\sum_{k=j}^{n-1} (-1)^{j+1} B^k P B^{n-k-1} \nonumber \\
&=&-\frac{1-(-1)^n}{2} B^n D +\sum_{k=0}^{n-1}\frac{1+(-1)^k}{2} B^k P B^{n-k-1}.
\end{eqnarray}
Inserting (\ref{eq11}) into (\ref{eq12}) and exchanging the sums results in
\begin{eqnarray}
\frac{\px}{\px s}(\Id+B)^{-1} &=& \sum_{n\geq 1} B^{2n+1} D +\sum_{k\geq 0}\sum_{n\geq k+1}\frac{1+(-1)^k}{2} B^k P (-B)^{n-(k+1)} \nonumber \\
&=& (\Id-B^2)^{-1} BD +(\Id-B^2)^{-1} P (\Id+B)^{-1}.
\end{eqnarray}
\end{proof}

\section{Outlook}
The asymptotic distribution of the largest eigenvalue is also known for Gaussian unitary ensemble of Hermitian matrices ($\beta=2$) and Gaussian symplectic ensemble of quaternionic symmetric matrices ($\beta=4$). As just established, for $\beta=1$,
\begin{equation}
F_1(s)=\det(\Id-B(s)),
\end{equation}
and, for $\beta=2$,
\begin{equation}
F_2(s)=\det(\Id-B(s)^2),
\end{equation}
which might indicate that $F_4(s)$ equals $\det(\Id-B(s)^4)$. This is however incorrect, since the decay of $\det(\Id-B(s)^4)$ for large $s$ is too rapid. Rather one has
\begin{equation}\label{eq13}
F_4(s/\sqrt{2})=\frac12\big(\det(\Id-B(s))+\det(\Id+B(s))\big).
\end{equation}
This last identity is obtained as follows. Let $U(s)=\frac12 \int_s^\infty q(x)\dx s$ with $q$ the unique solution of the Painlev\'e II equation $q''=sq+2q^3$ with $q(s)\sim \Ai(s)$ as $s\to\infty$. Then the Tracy-Widom distributions for $\beta=1$ and $\beta=4$ are given by
\begin{equation}
F_1(s)=\exp(-U(s)) F_2(s)^{1/2},\quad F_4(s/\sqrt{2})=\cosh(U(s)) F_2(s)^{1/2},
\end{equation}
see~\cite{TW96}. Thus $F_4(s/\sqrt{2})=\frac12 (F_1(s)+F_2(s)/F_1(s))$, from which (\ref{eq13}) is deduced.


\end{document}